\documentclass[final]{aipproc}

\layoutstyle{6x9}
\pdfoutput=1

\begin{document}

\title{Learning about Galactic structure with Gaia astrometry}



\author{Anthony G.A.\ Brown}{
address={Sterrewacht Leiden, Leiden University, P.O.\ Box 9513, 2300 RA Leiden,
The Netherlands}}

\begin{abstract}
  The Gaia mission is reviewed together with the expected contents of the final
  catalogue. It is then argued that the ultimate goal of Galactic structure
  studies with Gaia astrometry should be to build a dynamical model of our
  galaxy which is capable of explaining the contents of the Gaia catalogue. This
  will be possible only by comparing predicted catalogue data to Gaia's actual
  measurements. To complement this approach the Gaia catalogue should be used to
  recalibrate photometric distance and abundance indicators across the
  HR-diagram in order to overcome the lack of precise parallax data at the faint
  end of the astrometric survey. Using complementary photometric and
  spectroscopic data from other surveys will be essential in this respect.
\end{abstract}

\maketitle

\section{Overview of the Gaia mission}

Gaia will provide a stereoscopic census of our Galaxy through the measurement of
high accuracy (global and absolute) astrometry, radial velocities and
multi-colour photometry. Gaia will measure parallaxes and proper motions for
every object in the sky brighter than magnitude 20 --- amounting to over 1
billion stars, galaxies, quasars and solar system objects. It will achieve a
complete all-sky survey to its limiting magnitude via real-time on board
detection. The astrometric accuracy will be 12--25 $\mu$as, depending on colour,
at 15th magnitude and 100--300 $\mu$as at 20th magnitude. Multi-colour
photometry will be obtained for all objects by means of low-resolution
spectrophotometry. The photometric instrument consists of two 
prisms dispersing all the light entering the field of view. One disperser ---
called BP for Blue Photometer --- operates in the wavelength range 330--680 nm;
the other --- called RP for Red Photometer --- covers the wavelength range
640--1050 nm. In addition radial velocities with a precision of 1--15 km/s will
be measured for all objects to 17th magnitude, thus complementing the astrometry
to provide full six-dimensional phase space information for the brighter
sources. The radial velocity instrument (RVS) is a near-infrared (847--874 nm,
$\lambda/\Delta\lambda\sim 11\,000$) integral-field spectrograph dispersing all
the light entering the field of view. Gaia builds on the proven principles of
the Hipparcos mission but represents an improvement of several orders of
magnitude in terms of numbers of objects, accuracy and limiting magnitude
(Hipparcos observed $120\,000$ stars to 12th magnitude, achieving
milli-arcsecond accuracy).

The Gaia satellite and mission operations are fully funded by the European Space
Agency and the launch is foreseen in December 2011. Gaia will orbit the
Sun-Earth L2 Lagrange point at $1.5$ million km from the Earth. The transfer to
this point and the commissioning of the spacecraft and payload will take up to 6
months, after which the routine science operations start. This phase will last 5
years, with a possible 1 year extension. The data processing will be undertaken
by the scientific community in Europe which has organized itself into the Gaia
Data Processing and Analysis Consortium (DPAC). The processing will be ongoing
during the operational phase and for 2--3 years after. The final catalogue is
expected around 2021, but with intermediate data releases produced during the
operational phase.

In the context of Galactic structure studies it is interesting to consider the
information which the final Gaia catalogue will contain on stars. In Gaia's own
broad-band magnitude $G$ the number of stars will be $\sim 7\times10^5$ to
$G=10$, $48\times10^6$ to $G=15$ and $1.1\times10^9$ to $G=20$. About $60$
million stars are expected to be seen as binary or multiple systems by Gaia. For
each star the following information is provided:
\begin{description}
  \item[astrometry] positions, parallax, proper motions, the full covariance
    matrix of the astrometric parameters (standard errors and correlations) and
    astrometric solution quality indicators.
  \item[photometry] broad band fluxes in the $G$, $G_\mathrm{BP}$,
    $G_\mathrm{RP}$ and $G_\mathrm{RVS}$ bands, as well as the prism spectra
    measured by the blue and red photometers. Variability indicators will be
    provided for all stars together with epoch photometry.
  \item[spectroscopy] the radial velocity is listed for each star observed by
    RVS and for the brighter stars $v\sin i$ as well as the accumulated spectra
    will be provided.
  \item[multiple stars] solution classifications and, where relevant, orbital
    parameters together with covariance matrices and quality flags.
  \item[astrophysical parameters] the Gaia catalogue will provide as much
    astrophysical information on each star as possible, derived from the
    combination of photometric, spectroscopic and parallax information. The
    astrophysical parameters include $T_\mathrm{eff}$, $A_V$, $\log g$, [M/H],
    and [$\alpha$/Fe] where possible. Luminosities, ages, and variable star
    characterizations will also be provided.
\end{description}
The catalogue will also contain intermediate data which allow a reprocessing of
the observations. In particular the astrophysical parameters can be improved by
incorporating complementary information obtained from other surveys or follow-up
observations. For astrometric data reprocessing is relevant for multiple sources
with uncertain solutions (which can be improved by incorporating follow-up
radial velocity observations) or, for example, in the case of proper motion
measurements for faint stars in a dwarf galaxy orbiting the Milky Way one could
combine the measurements for all stars to derive a (more precise) mean motion
and parallax for the dwarf. Hence the Gaia catalogue will not be a static
release but will evolve over time as more information is added to it.

For much more information on the Gaia mission and the science topics that will
be addressed I refer to the proceedings of the symposium \emph{The
Three-Dimensional Universe with Gaia} \cite{Turon2005} and recent overviews of
the mission status \cite{Lindegren2008} and DPAC \cite{Mignard2008}.

\section{Galactic structure studies with Gaia}

Before highlighting some of the Galactic structure topics where Gaia's
astrometric data is expected to have a major impact it is useful to consider its
astrometric performance in relation to other large existing or upcoming surveys
of the Galaxy (see \cite{Brown2008} and \cite{Turon2008}). From this comparison
it is clear that Gaia will be the only mission in the coming decade that
provides astrometry in the $10$--$100$~$\mu$as regime in the optical for large
all-sky samples and covering a large volume throughout our Galaxy. Important
complementary astrometry will be provided by the JASMINE mission in the near
infrared (covering the bulge and inner disk regions), the SIM Planetquest
mission in the optical to accuracies of a few $\mu$as for selected targets, and
by the LSST and Pan-STARRS surveys which will provide milli-arcsecond astrometry
to 24th magnitude.

Some of the top questions concerning our galaxy are listed in \cite{Turon2008}.
What is the mass distribution throughout the Galaxy? What is the spiral arm
structure?  What is the merging history of our galaxy? How many mergers occurred
and what did the building blocks look like? Is our galaxy consistent with the
$\Lambda$CDM paradigm? How and when were the structural components (bulge, halo,
thin/thick disk) of the Galaxy assembled? Which stars formed where and when did
they form?  The most direct impact of Gaia astrometry on these questions will be
through the vast amount of stars for which accurate 5-dimensional phase space
data will be available. Indirectly the astrometric data, in particular the
parallaxes, will enable a highly accurate mapping of the Hertzsprung-Russell
diagram, thus tremendously improving our understanding of stars and their
atmospheres. As a consequence the astrophysical characterization of stars will
be improved leading to better age and chemical composition determinations. This
enables a detailed reconstruction of the evolution of the Galaxy and its
structural components in time. A few examples of Galactic structure studies with
Gaia are discussed in some more detail below. Much more examples and extensive
discussions can be found in \cite{Turon2008}.

In a large volume around the Sun the space motions of stars will be accurately
determined (using the RVS data) which allows the reconstruction of their orbits
and the tracing of their birthplaces in the Galaxy. When complemented by precise
stellar ages and the possibility to construct volume limited samples for
spectroscopic follow-up studies, the detailed reconstruction of the history of
the disc becomes possible by disentangling the effects of stars formation
history, dynamics, and chemical evolution of the interstellar medium on the
observed properties of stars.

The stellar halo was largely built up through accretion and/or mergers and a
major goal is to identify the building blocks. Recently many tidal streams and
dwarf satellites orbiting our Galaxy have been identified, in large part due to
the Sloan Digital Sky Survey. However these objects are mostly confined to the
outer halo and the analysis of stellar abundances for dwarf satellites and field
halo stars make it clear that these newly discovered streams and dwarfs are not
the building blocks of the bulk of the halo. The accretion history of the halo
is thus to a large extent contained in its field stars and tracing these to the
original building blocks is difficult as they have been thoroughly mixed in
configuration space. Gaia is uniquely capable of providing the necessary phase
space information for the inner halo which is needed in combination with the
astrophysical information on the stars in order to reconstruct the formation
history of the halo.

Finally, Gaia will tremendously advance the mapping of the spiral structure of
our galaxy. A direct tracing of spiral arms is possible through the accurate
distance information for early type stars. Indirectly the arms can be traced
through the stellar kinematics in the disk. Gaia parallaxes are also required
for the construction of a 3D extinction map which is needed to overcome the
obscuring effects of dust in the plane of the Milky Way.

\section{From Gaia to a dynamical model of the Galaxy}

Studies of specific Galactic components are important but do not provide the
whole picture. The structural components are coupled through gravity and the
observed stellar and gas kinematics are determined by the gravitational
potential of the Galaxy. The only way to develop a consistent understanding of
the mass distribution and kinematics is through a dynamical model of the Galaxy,
and it is only with such a model that one can make reliable extrapolations to
the unobserved parts of Galactic phase space. The Gaia catalogue can be seen as
a snapshot of the state of the Galaxy in which we will be seeing stars from the
same population at different points along the same orbits. This allows the
reconstruction of individual orbits from which we can infer the Galactic
potential and matter distribution. Any dynamical model will thus be highly
constrained.

Hence as argued in \cite{Binney2005}, if we want to take full advantage of an
all-sky high accuracy astrometric data-set (complemented by radial velocities,
photometry and astrophysical information) and convert this data for 1 billion
stars into a complete physical understanding of the structure of our galaxy, the
goal should really be to construct a dynamical model in terms of which we can
explain the entire Gaia catalogue. This is obviously a non-trivial task and I
discuss below the steps that should be taken in anticipation of the Gaia
catalogue and once it is available.

\subsection{Dynamical modelling techniques}

Constructing such a dynamical model is a difficult task as it has to be able to
self-consistently determine matter and velocity distributions from the
underlying potential. Moreover, in comparing with the Gaia catalogue data the
astrophysical properties of the stellar populations (composition, age) as well
as the effects of extinction due to dust have to be accounted for. A dedicated
programme to prepare such models is therefore needed as described in
\cite{Binney2005}. The latter paper describes three promising approaches to the
construction of the required Galaxy model; extensions of the Schwarzschild
method, made-to-measure N-body techniques, and the torus technique.

The Schwarzschild method \cite{Schwarzschild1979} has been applied extensively
to external galaxies in order to derive their phase space structure from
observations of the line of sight velocity distributions or integral field
spectrograph data. The method works by calculating an orbit library in a trial
potential and then finding the weighted superposition of orbits which can
reproduce continuous observables such as surface brightness and velocity
moments. The application to Gaia data requires extending the method so it can be
used with discrete data for individual stars. A first step in this direction has
been taken in \cite{Chaname2008} but the method will have to be extended to
triaxial geometries and has to be capable of dealing with effects such as that
of the Milky Way's bar.

The made-to-measure N-body method was introduced in \cite{Syer1996} and applied
to the Milky Way as described in \cite{Bissantz2004}. Unlike the Schwarzschild
method this technique does not pre-calculate large libraries of orbits followed
by a weighted superposition. Rather, the weights are determined simultaneously
with the integration of the orbits. This method can be incorporated into N-body
simulations (e.g., to generate equilibrium initial conditions for large-$N$
systems) and is flexible in that it allows for arbitrary geometries and
potentials that evolve in time. The challenge will be to overcome the
computational complexity involved when trying to reproduce the Gaia catalogue
data.

Finally, the torus technique \cite{Dehnen1996} replaces the orbit library used
in the Schwarzschild method with phase-space tori. This method requires complex
software and has so far only been demonstrated for axisymmetric systems. In
addition the orbits are derived in an integrable potential while the true
potential of the Milky Way is certainly not integrable. The torus method
requires extensive development in order to be able to deal with realistic Milky
Way potentials. Some ideas are given in \cite{Binney2005}.

The development of these techniques should start now but we do not have to wait
for the Gaia catalogue to be finished before applying them, as there will be
plenty of surveys providing photometric and kinematic data for large samples of
stars in our galaxy.

\subsection{Finding the best Galaxy model}

As noted in \cite{Binney2005} any of the methods just described will lead to a
dynamical model of the Galaxy consisting of a gravitational potential and
distribution functions for each of the stellar populations. The latter are
described as probability distributions in mass, chemical composition and age for
the stars in a particular population. Hence the Galaxy model will consist of a
very large number of parameters and `fitting' the model to the entire Gaia
catalogue will be a very challenging task.

The basic predictions from the models are the distributions of the stars in
phase-space $(\mathbf{r},\mathbf{v})$ at some time. These distributions could be
compared to phase-space variables calculated from the Gaia astrometric and
radial velocity data in order to decide on the best Galaxy model. This approach
suffers from the fact that for most stars the radial velocity will not be
available, but more importantly it requires the transformation from observed
parallaxes to distances. It is important to keep in mind that we do {\em not}
directly measure the distance to stars. What we measure instead are their
parallactic displacements on the sky caused by the motion of the earth around
the sun. The transformation of the observed parallax $\varpi_o$ to distance,
given by $r_o=1/\varpi_o$, is seemingly trivial but can cause many problems
because it is non-linear. In the absence of systematic measurement errors the
observed parallax is itself an unbiased estimate of the true parallax $\varpi$:
$E[\varpi_o]=\varpi$. This is not true of the distance, i.e.\
$E[1/\varpi_o]\neq1/\varpi$. This well-known problem was discussed at the time
of the release of the Hipparcos Catalogue \cite{Brown1997} and plays an
important role in luminosity calibrations where effects of the Lutz-Kelker type
can lead to biased estimates of the absolute magnitudes of stars. The same
problems occur when calculating the positions and motions of stars (all
proportional to $1/\varpi$) or integrals of motion such as energy $E$ or angular
momentum $L$ (both functions of $1/\varpi^2$). The energy-angular momentum plane
is is a powerful tool when looking for remnants of accreted satellites. However
as shown in \cite{Brown2005} the propagation of parallax errors can lead to sign
changes in $L_z$, spurious features in the $E$-$L_z$ plane, and in addition the
$1/\varpi^2$ dependence will cause the estimates of the integrals of motion for
a particular satellite to be biased.

The only truly robust way to get around this problem is to project the Galaxy
model into the data-space and thus predict the astrometric data together with
the other data in the Gaia catalogue (radial velocities, magnitudes and colours
of stars). The added advantage is that one can easily account for incomplete
phase space data (e.g., lack of radial velocity data) and selection effects. The
extinction due to dust can be taken into account in predicting the observed
distribution of magnitudes and colours of the stars. Moreover, negative
parallaxes (which are perfectly legitimate measurements!) and the correlations
in the errors on the astrometric parameters (which will vary systematically over
the sky) can be much more easily accounted for in the data-space.

To decide on the best values for the Galaxy model parameters one would ideally
use the maximum likelihood technique. However, given the complications of the
Galaxy model and the large number of parameters it will very likely not be
possible to construct the likelihood function and its derivatives (need for its
maximization). One will therefore have to resort to generating mock Gaia
catalogues from the model and comparing these to the actual data. The challenges
here are the comparison of predicted and observed distributions of observables
for very large amounts of data and the exploration of a very high dimensional
model space in order to find the optimum parameter values. The result should be
a probability distribution over the model space (where we should keep in mind
that not all aspects of the Galaxy model will be uniquely determined). The
problem of finding this distribution for a very high dimensional parameter space
is a well known issue in inverse problem theory \cite{Tarantola2005}. Many
efficient methods for exploring the parameter space have been developed over the
years, such as Markov Chain Monte Carlo methods, and our quest for a dynamical
model of the Milky Way can take advantage of this knowledge. An extensive
discussion on how the Torus technique can be combined with the inverse problem
methodology is given in \cite{Kaasalainen2008} for the case of spherical
systems.

\subsection{Gaia alone is not enough}

To get the dynamical modelling effort started a direct examination of the
positions and kinematics of the stars will still be essential. This will enable
us to characterize the structural components of the Galaxy, the degree of non
axisymmetry, and the amount and kind of substructure present in the halo. The
identification and description of Galactic stellar populations will provide
further constraints on the model (see \cite{Kaasalainen2008}) and is essential
to the study of Galactic structure as a function of time.

In order to do this accurate alternative distance indicators are needed for the
faint and distant stars in the Gaia survey to supplant their relatively
imprecise parallaxes. This will also enhance the separation and description of
stellar populations through better determined luminosities, ages, and chemical
compositions. Gaia will enable us to recalibrate to high accuracy photometric
distance indicators across the HR-diagram. To illustrate this
Table~\ref{tab:reach} shows Gaia's reach in terms of relative parallax precision
for a selected set of stellar types.

\begin{table}
  \begin{tabular}{lccccccccc}
    \hline
    & \multicolumn{4}{c}{No extinction ($A_V=0$)} & & \multicolumn{4}{c}{$A_V=5$
    mag} \\
    \cline{2-5} \cline{7-10}\\
    Type of star & 1\% & 2\% & 5\% & 10\% & & 1\% & 2\% & 5\% & 10\% \\
    \hline
    G0V ($M_V=+4.4$) & $0.8$ & $1.1$ & $1.8$ & $2.5$ & & $0.3$ & $0.5$ & $0.7$ &
    $1.0$ \\
    K5III ($M_V=-0.1$) & $1.3$ & $2.6$ & $4$ & $7.5$ & & $1.0$ & $1.5$ & $2.4$ &
    $3.5$ \\
    Cepheid ($P=10^\mathrm{d}$, $M_V=-4.1$) & $1.2$ & $2.4$ & $6$ & $12$ & &
    $1.2$ & $2.3$ & $3.8$ & $7$ \\
    \hline
  \end{tabular}
  \caption{Parallax horizon of Gaia. This is the maximum distance (in kpc) at
  which a given relative precision in parallax (or distance) is obtained. Based
  on latest accuracy estimates, with $M_V$ from \cite{Cox2000}. Table reproduced
  from \cite{Lindegren2008}.\label{tab:reach}}
\end{table}

Precise photometric distance indicators cannot be derived from the Gaia data
alone as at the faint end the determination of extinctions, surface gravities
and abundances from Gaia's photometry will not be reliable. Complementary data
from other large area surveys such as Pan-STARRS, LSST, and SkyMapper, as well
as follow-up spectroscopic observations, are needed to calibrate photometric
abundance and $\log g$ indicators. Wide-field multi-object spectrographs
covering the UV/blue, deployed on 4 and 8 meter class telescopes, will be
essential in this respect. The complementary photometric data in combination
with data at longer wavelengths is also needed to determine precise extinction
values for stars in the Gaia catalogue.

\smallskip
In conclusion, the Gaia mission will provide a vast and highly accurate
astrometric data-set, complemented by astrophysical information, which offers a
fantastic opportunity to obtain a complete physical understanding of our Galaxy.
We should start preparing now for the challenge of building the required
dynamical model.

\end{document}